# A Comprehensive Analysis of Twitter Trending Topics

Issa Annamoradnejad
Department of Computer Engineering
Sharif University of Technology
Tehran, Iran
i.moradnejad@gmail.com

Jafar Habibi
Department of Computer Engineering
Sharif University of Technology
Tehran, Iran
jhabibi@sharif.edu

*Abstract*— In Twitter, a name, phrase, or topic that is mentioned at a greater rate than others is called a "trending topic" or simply "trend". Twitter trends list has a powerful ability to promote public events such as natural events, political scandals, market changes and other types of breaking news. Nevertheless, there have been very few works focused on the dynamics of these trending topics. In this article, we thoroughly examined the Twitter's trending topics of 2018. To this end, we automatically accessed Twitter's trends API and stored the resulting 50 top trending topics in a novel dataset. We propose and analyze our dataset according to six criteria: lexical analysis, time to reach, trend reoccurrence, trending time, tweets count, and language analysis. Based on our results, 77.6% of the topics that reached the Top-10 list were trending with less than 100k tweets. More than 50% of the topics could not hold the position for more than an hour. English and Arabic languages comprised close to 40% and 20% of the first rank topics, respectively.

I. INTRODUCTION

In the last two decades, online social media sites (such as Facebook, Twitter, Youtube, etc.) have revolutionized the way we communicate with each and transformed everyday practices [1]. Among these Online Social Networks (OSNs), Twitter is a free long range microblogging website that enables enlisted individuals to communicate short posts called tweets. Twitter users can communicate tweets and take after other client's tweets by utilizing numerous activities [2]. Twitter is currently the third most popular social networking service, with more than 335 million active users [3].

A name, phrase, or topic that is mentioned at a greater rate than others is called a "trending topic" or simply "trend". Trending topics become popular either through a concerted effort by users (as in promoting an election nominee) or because of an event that prompts people to talk about a specific topic (such as a TV series or earthquake). A list of top ten trending topics is listed in the website, which help Twitter and their users to understand what is happening in the world and what people's opinions are about it.

Twitter trends has shown their powerful ability in many public events, such as in the wildfires in San Diego and the earthquake in Japan [4]. In addition, governments and businesses analyze and understand the dynamics of general mood of population to reach better results. Some previous works studied the importance of Twitter in detecting real time events, predicting market fluctuations and even election results. Nevertheless, there has been very few works that focused on understanding the dynamics and statistics of these trending topics.

In this article, we thoroughly examined the Twitter's trending topics of 2018. To this end, we accessed Twitter's trends API for the full year of 2018 and generated a full dataset. Contrary to the top ten list shown in the website, the API returns a list of top 50 trending topics for a given place. A version of our dataset (hourly basis) is provided at [5].

To analyze this aggregated dataset in several aspects, we devised six criteria, which are: lexical analysis, time to reach, trend reoccurrence, trending time, tweets count, and language. We examined our dataset in these six criteria according to three conditions: First rank trends, Top10 and Top50 list. In addition to computing general statistics about each criterion, determining longest trending topics, topics with highest tweet counts, most reoccurring topics, and most used languages, we will generate related distributions such as: tweets count distribution, reoccurrence count distribution, trending time distribution, language distribution, and words and characters count distributions.

The structure of this article is as follows: Section 2 reviews past works on Twitter trends and related topics, alongside a brief description of used terms in Twitter. Section 3 explains the data and methodology. In Section 5, results and discussion for the six criteria are given, and Section 6 is the concluding remarks. Background

*A. Literature review*

Many examples from the real world events have demonstrated the effectiveness of trending topics in attracting more attention from the world, during disasters and social movements, and there have been good body of research to provide analytics for those events. Reference [4] study social, spatial, and temporal characteristics of earthquake-related tweets, [6] describe a method for using Twitter to track forest fires and the response to the fires by Twitter users, [7] analyzed information diffusion activity during the 2011 Egyptian political uprisings, [8] examined the tweets and strategies corresponding to the United States Presidential election, [9] used spike detection algorithm to detect the important moments within sporting events like World Cup Soccer matches, which take place over a short period of time, and [10] analyzed the topical, geographic, and temporal importance of descriptors for three events over time that can help visualize the event data.

In contrast, very little study has been done to explain or understand Twitter trends dynamics. Reference [11] classified trends into 18 categories using two separate approaches. Their results based on 768 unique trending topics, showed that Sports, Music and Movies had the highest numbers of trending topics, respectively. Reference [12] proposed Sequential Summarization to generate a serial of chronologically related sub-summaries for a given topic while retaining the order of information presentation and each sub-summary attempts to concentrate on one theme or subtopic. Some researchers (such as [13], [14]) attempted to aggregate several explanations into one long summary using traditional summarization approaches, but it still loses much useful information, such as the change of Twitters' focus and the temporal information. Reference [15] provided an evaluation on the methods of trends disambiguation to find the most successful method that uses to retrieve the representative contents of trending topics.

In addition, some services have been dedicated to explain Twitter trends autonomously or in a collaborative scenario (more information in [12]). Applications like echofon (www.echofon.com), whatthetrend (whatthetrend.com) have evolved from Twitter, which provide services to explain why a term becomes a trending topic or to give a short description of the trending topic. These applications or services generally track the topics in Twitter to automatically generate a description for a given topic. In some cases, they encourage users to give small summaries on a new tweet to explain the topics. For example, whatthetrend encourages users to edit explanatory tweets about topics and ranks the submitted explanatory tweets by readers' agreements, providing a good way to help users understand the topic.

### B. Twitter Service Overview

Twitter is currently the eleventh most popular site in the world according to the Alexa traffic rankings [3]. This popularity is also in the large number of research papers published about Twitter in various fields, such as social network analysis or data mining. Twitter's core function allows users to share any kind of information, thoughts, opinions, and ideas to keep people updated or informed of happenings short messages, in several formats including text, images and videos. The social relationship on Twitter is asymmetric and can be conceptualized as a directed social network or follower.

Users are encouraged to categorize their tweets by a hashtag, which is any keyword preceded by a hash sign "#" (e.g., #sorcery). This allows users for faster content discovery or to track specific events in real time. Twitter displays top ten trending topics for several places as well as worldwide trends, which over the years has gained a lot of attention and news coverage.

Twitter provides an application programming interface (API), which allows developers to programmatically access the public data streams, search the old tweets, access trending topics and many other features of the service. The availability of Twitter data has motivated significant research work in various disciplines and led to numerous applications and tools.

## II. DATA AND METHOD

In this section, we will briefly explain our data gathering method, alongside some general statistics on our collected dataset. In addition, we will explain the methods we will utilize in the next chapter to analyze our data.

### A. Data

We collected our dataset using the Twitter's trends API[1], which provides current trending topics for a given place in the format of WOEID[2]. Contrary to the trending list in the website which displays only 10 items, the API result consists of (mostly) 50 trending topics with their corresponding tweets' count (where available). We accessed the API for worldwide trending topics using a script for the entire year of 2018. The script accessed the API every 10 minutes with 97.35% availability. It should be noted that because of the script time interval, there could be a few uncovered states, such as those trends that held the first rank for less than 10 minutes. A version of our dataset (hourly basis) is provided at [5].

Our dataset is comprised of 155899 unique topics which were listed once or more in the top50 trending topics. 68% of the topics were hashtags, while the rest of them were names or expressions without the hashtag sign (#). Tweets count was available for more than 96% of the first rank topics, 72% of the topics in the top10 list and 43% of all aggregated items.

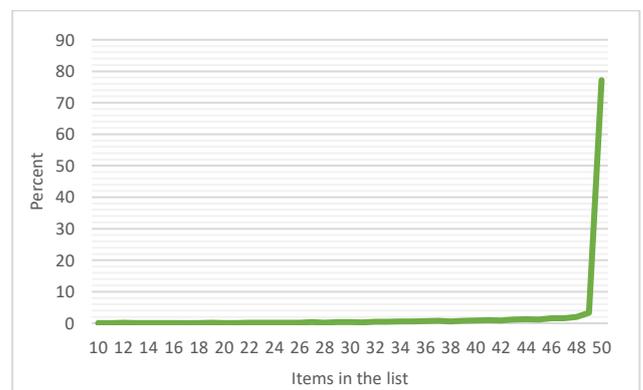

Fig. 1. Distribution of the count of available trending topics provided by the API

---

[1] Can be accessed from https://api.Twitter.com/1.1/trends/place.json

[2] Where On Earth IDentifier

As we briefly mentioned earlier, not all of our requests to the API resulted in a list of 50 items. Fig. 1 displays the distribution of the count of available trending topics provided by the API. As it is clear, the API provided the full list of 50 topics in most cases (77.1%).

*B. Method*

We will analyze our dataset with respect to six criteria:

1. *Lexical analysis* which examines trending topics in a lexical level; such as the number of words or letters used in the trend.

2. *Time to reach* which calculates the time it takes for a topic at the bottom of the list to reach to higher positions, such as the first rank or the top10 list.

3. *Trend reoccurrence* which counts the number of times a topic will become trending. For example, this happens weekly for hashtags related to a TV series.

4. *Trending time* which scores trends based on the time they were trending non-stop. Any of these *trending times* is considered one reoccurrence (the previous criteria).

5. *Tweets count* which examines trending topics based on the number of tweets related to each one. As we discussed this in previous section, this number was available in 43% of all cases.

6. *Language* which analyzes trending topics based on their language. To this end, first we stripped texts from any underlines or # symbols to get a clean expression clear. We also converted camelCased hashtags to normal spaced expressions. Then, using *textblob* library for *python*, we detected language for all of the trending topics.

In addition to evaluating some general statistics about each one of the six criteria and exploring the top items based on some of them, we computed several related distribution charts which give insights into this bulk of data.

We will perform our analyses in three conditions:

1. *First rank* which focuses on the top trending topic and discards the rest of the list as being a trending topic. For example, if a topic goes down from the first rank to the second rank and then goes up, it is considered a reoccurring event.

2. *Top10* which focuses on the top 10 trending topics. This list can be accessed from the main website.

3. *Top50* which analyzes the complete list of trending topics provided by the API.

### III. RESULTS

In this section, we present our results of analyses based on the aforementioned six criteria on the trending topics of 2018. Results for each criterion are provided in a separate subsection.

*A. Lexical Analysis*

Based on our analysis on our dataset, more than half of the trending topics were composed of a single word. *#HowToGetAwayWithScandal* and *#EiBozonaroVaiTomarNoCu*, which respectively became the most trending topics in March 2$^{nd}$ and September 22$^{nd}$, used the highest number of words among the first rank topics of 2018.

In addition to the analysis of word count, we examined the number of characters used in trending topics. Character count ranges between 2 and 31 characters. While there was only one topic with two characters (*Eu*), six unique first rank topics used the highest number of characters and interestingly, all were in Arabic language.

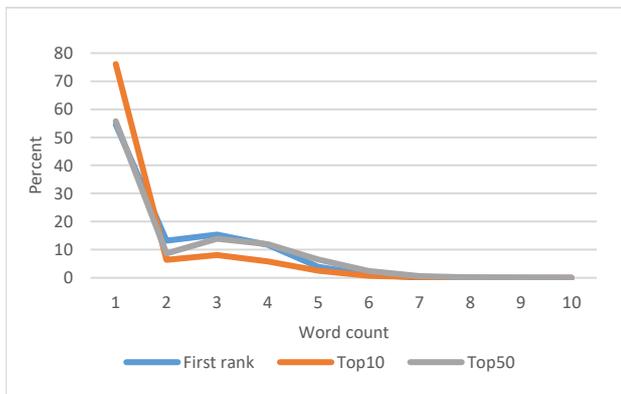

Fig. 2. Distribution of the number of words used in trending topics

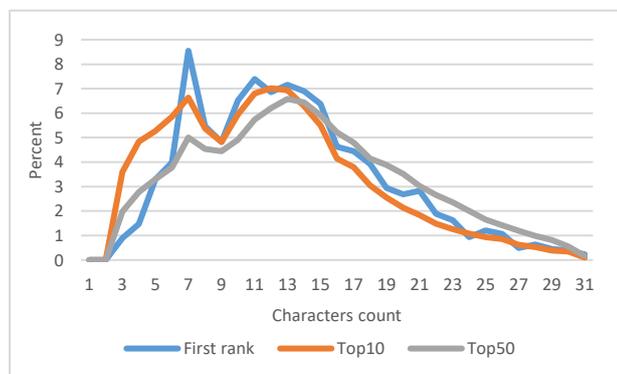

Fig. 3. Distribution of the number of characters used in trending topics

Fig. 2 and Fig. 3 display the distribution of the number of words and characters used in the trending topics for the three conditions of the previous chapter, respectively. In average, trending topics were composed of around 13 characters and 2 words.

*B. Time to reach*

*Time to reach* calculates the minutes it takes for a topic to reach from the bottom of the list to higher positions. We analyzed it in two conditions: Time to reach to the first rank and time to reach to the top10 list. Fig. 4 represents the distributions of time to reach for the two conditions.

In average, it takes 36.2 minutes for a topic to reach the top10 list and 91.5 minutes to reach the first rank. It goes without saying that not all trending topics can reach the first rank position. There were 977 topics that reached the first rank in less than 10 minutes which comprises more than 28% of all first rank topics. And, there were more than 19 thousand topics that reached the top10 list in less than 10 minutes, which is more than 48% of all top10 items.

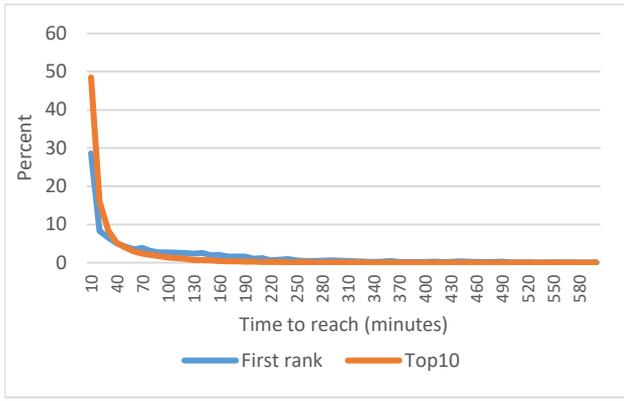

Fig. 4. Distribution of the time to reach to higher ranks

## C. Trend Reoccurrence

*Trend Reoccurrence* happens when a topic reaches high trending ranks, then goes off the chart and comes back later. In average, a topic reoccurred less than 1.5 times, meaning that most trending topics were trending for only one time in the whole year. The most reoccurring trend in the first rank spot is *#FelizLunes*, which held the position for 46 separate times. *#WednesdayWisdom* was the most recurring trend in the top10 chart (95 times), and *#FlashbackFriday* was the most recurring trend in the top50 chart (144 times). Ten highest reoccurring trending topics is available in Table I.

As it is clear in Table I, all of the items except one are hashtags. More than 86% of the first rank topics only show up once in that position, and only 2% of them reoccurred more than 3 times. Fig. 5 represents the distribution results of reoccurrence for the three conditions.

## D. Trending time

In this subsection, we will examine the time of trending without any break. In average, a topic stays in the first rank position for 99.3 minutes, in the top10 list for 85.6 minutes and in the top50 chart for 101.9 minutes. This is an interesting result as it means that although there are more competition to reach the first rank (since it has only one spot to fill), the winner holds the position for longer time compared to other spots of the top10 list.

Table II shows the top trending topics according to their trending time in the three conditions of the past chapter. *#InternationalCatDay* held the first rank position for close to 13 hours in Aug-8, #الفيصلي_الاتحاد was in the top10 list for more than 18 hours and in the Top50 chart for more than 21 hours in May-12.

More than 17% of the topics held the first rank position for less than 10 minutes, and more than 50% of the topics could not hold the position for more than an hour. Fig. 6 represents the distribution of trending time in the three conditions. It should be noted that almost 10% of the trending topics, were trending for more than 4 hours, which are omitted from this figure.

TABLE I. TOP TEN REOCCURRING TRENDING TOPICS OF 2018

| | First rank | # | Top10 | # | Top50 | # |
|---|---|---|---|---|---|---|
| 1 | #FelizLunes | 46 | #WednesdayWisdom | 95 | #FlashbackFriday | 144 |
| 2 | #FelizMartes | 40 | #FelizMiércoles | 88 | #QuintaDetremuraSDV | 116 |
| 3 | #FelizJueves | 35 | #TuesdayThoughts | 74 | #SabadoDetremuraSDV | 114 |
| 4 | #MasterChefBR | 27 | #FelizDomingo | 73 | #QuartaDetremuraSDV | 106 |
| 5 | #BuenViernes | 21 | #ThursdayThoughts | 71 | #SegundaDetremuraSDV | 105 |
| 6 | #MondayMotivation | 21 | #FelizJueves | 69 | #TercaDetremuraSDV | 102 |
| 7 | #FelizMiércoles | 20 | #FelizSábado | 69 | #DomingoDetremuraSDV | 101 |
| 8 | #WednesdayWisdom | 13 | #FelizMartes | 65 | #SextaDetremuraSDV | 99 |
| 9 | #FelizDomingo | 12 | #BuenViernes | 61 | #BuenMiercoles | 93 |
| 10 | Messi | 11 | #FelizLunes | 53 | #ThursdayThoughts | 83 |

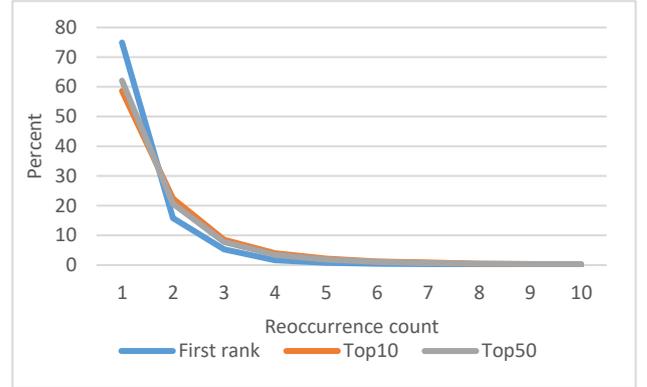

Fig. 5. Distribution of the number of trend reoccurrence

TABLE II. TOP TEN TOPICS BASED ON THEIR TRENDING TIME

| | First rank | t (H) | Top10 | t (H) | Top50 | t (H) |
|---|---|---|---|---|---|---|
| 1 | #InternationalCatDay | 12.7 | #الفيصلي_الاتحاد | 18.5 | #الفيصلي_الاتحاد | 21.3 |
| 2 | Thousand Oaks | 12.3 | #Caixa2doBolsonaro | 15.7 | مصر_روسيا | 20.2 |
| 3 | #كلنا_ثقه_في_محمد_بن_سلمان | 11.5 | #NationalDogDay | 14.8 | #WhyIJoinedTwitter | 19.8 |
| 4 | #KavanaughHearings | 11.2 | #InternationalCatDay | 14.5 | #ايش_استفادوا_الناس_منك | 19.7 |
| 5 | #Caixa2doBolsonaro | 11.0 | #فاتوره_الكهرباء | 14.3 | #استقبال_احمد_موسي_الاسطوري | 19.3 |
| 6 | #WorldMentalHealthDay | 10.5 | Thousand Oaks | 14.2 | #10Kasım | 19.2 |
| 7 | Happy Thanksgiving | 10.2 | #السعوديه_الاورغواي | 14.0 | #WoolseyFire | 18.8 |
| 8 | #8YearsOfOneDirection | 9.8 | #FridayFeeling | 14.0 | #نجدد_العهد_لولي_العهد | 18.7 |
| 9 | #HappyHalloween | 9.8 | Asia Argento | 13.7 | #MyFavoriteThingAboutThe90s | 18.3 |
| 10 | #SeduceMeIn4Words | 9.5 | #AnswerNow | 13.7 | #دافوس_الصحراء | 18.3 |

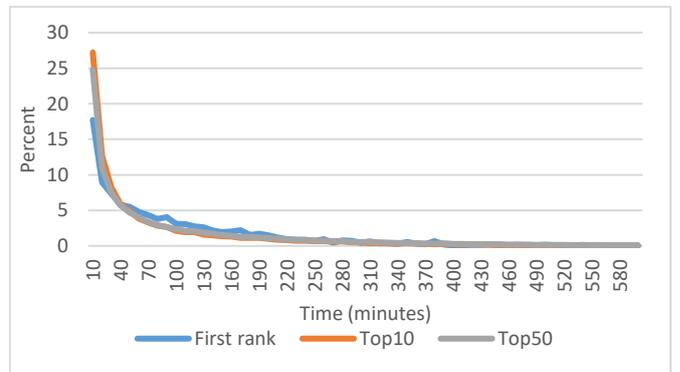

Fig. 6. Distribution of the trending time

*E. Tweets count*

As it was mentioned earlier, Twitter API provides the amount of tweets related to each of the trending topics. Based on our results, in average, a first rank topic has around 263,000, a topic in the top10 list has around 112,000 and a topic in the top50 chart has close to 73,000 related tweets. In addition, almost 55.9% of the topics that reached the first rank, 77.6% of the topics that reached the top10 list and 86.2% of all available trending topics (top50 chart), were trending with less than 100 thousand tweets.

Table III lists the top ten trending topics according to their tweets count. *#iVoteBTSBBMAs* which held the first rank position in May-14, had the highest amount of related tweets in 2018, and is reported to break the Guinness record [16]. Fig. 7 provide more insights into the distribution of trending topics according to their tweets count. It should be noted that the cases with more than 200,000 tweets were omitted from this figure for their negligible values.

*F. Languages*

Using *python* language and *textblob* library, we detected the written language for all of the trending topics. Based on our results, more than 81% of all trending topics are in ten languages. These ten languages alongside their corresponding percentages are represented in Fig. 8.

*English* language with 6,914 unique topics, comprised less than 40% of all topics. This is an interesting result, since the percentage of non-English trending topics has increased enormously compared to an analysis from 2011 [11] which stated that more than 87% of all trending topics were in *English* language. *Arabic* language is in the second place, by having

TABLE III. TOP TEN TRENDING TOPICS OF 2018 BASED ON THEIR RELATED TWEETS COUNT

|    | First rank | # | Top10 | # | Top50 | # |
|----|---|---|---|---|---|---|
| 1  | #iVoteBTSBBMAs | 16.0M | #iVoteBTSBBMAs | 17.5M | #iVoteBTSBBMAs | 17.5M |
| 2  | #MAMAVOTE | 15.7M | #MAMAVOTE | 15.7M | #MAMAVOTE | 15.7M |
| 3  | #ivotebtsbbmas | 13.3M | #BBMAs | 14.3M | #BBMAs | 14.3M |
| 4  | #BBMAs | 11.0M | #ivotebtsbbmas | 13.3M | #ivotebtsbbmas | 13.3M |
| 5  | #MAMARedCarpet | 10.9M | #MAMARedCarpet | 11.3M | #MAMARedCarpet | 11.3M |
| 6  | #TwitterBestFandom | 9.1M | #TwitterBestFandom | 9.5M | #TwitterBestFandom | 9.9M |
| 7  | #Olympics_EXO | 6.4M | #Olympics_EXO | 8.0M | #Olympics_EXO | 8.0M |
| 8  | Happy+New+Year | 6.4M | Happy+New+Year | 6.8M | Happy+New+Year | 6.8M |
| 9  | #TeenChoice | 6.3M | #TeenChoice | 6.7M | #TeenChoice | 6.7M |
| 10 | Stan+Lee | 4.4M | fake+love | 5.2M | fake+love | 5.2M |

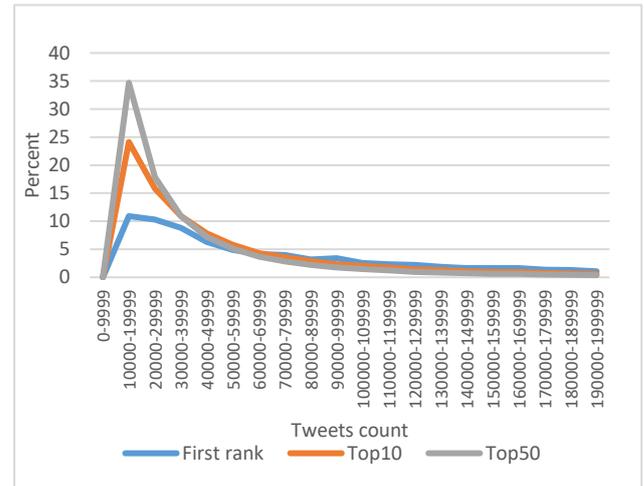

Fig. 7. Distribution of the number of related tweets

close to 20% of all first rank topics and 12% of all trending topics.

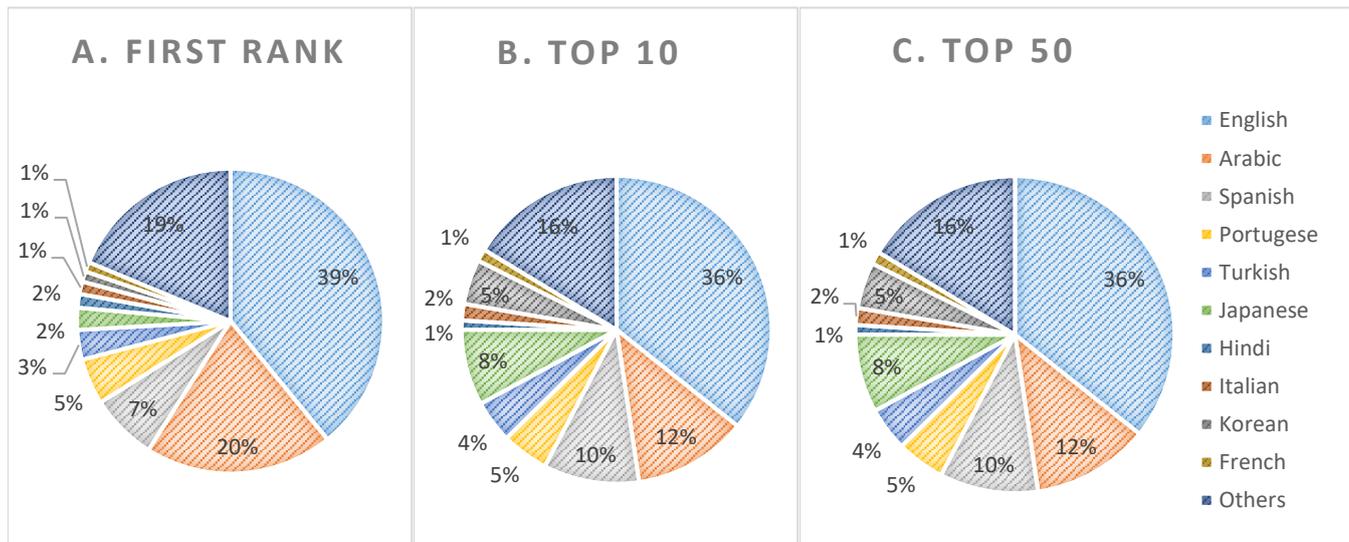

Fig. 8. Distribution of trending topics based on their language. Part A, B and C, display distribution by considering first ranked trends, Top10 list and Top50 chart, respectively.

## IV. CONCLUSION

In this article, we collected and analyzed Twitter trending topics of 2018. To examine these bulk of data in many aspects, we devised six criteria, which are: lexical analysis, time to reach, trend reoccurrence, trending time, tweets count, and language. We examined our dataset in these six criteria according to three conditions: First rank trends, Top10 and Top50 list.

In addition to computing general statistics in regard to each criterion, determining longest trending topics, topics with highest tweet counts, most reoccurring topics, and most used languages, we computed related distributions, such as: tweet count distribution, reoccurrence count distribution, trending duration distribution, language distribution, and words and characters count distributions.

Based on our results, more than 17% of the topics held the first rank position for less than 10 minutes, and more than 50% of the topics could not hold the position for more than an hour.

*English* and *Arabic* languages comprised close to 40% and 20% of the first rank topics, respectively. This is an interesting result, since the percentage of non-English trending topics has increased enormously compared to 2011 when more than 87% of all trending topics were in *English* language.

## REFERENCES

[1] D. M. Boyd and N. B. Ellison, "Social network sites: Definition, history, and scholarship," *J. Comput. Commun.*, 2007.

[2] T. Samakoti, S. Naseera, and T. Pramila, "Perception of trend topic in Twitter: A case study," *i-manager's J. Softw. Eng.*, vol. 11, no. 4, pp. 12–17, 2017.

[3] Alexa, "Alexa Top Sites," 2019.

[4] T. Sakaki, M. Okazaki, and Y. Matsuo, "Earthquake shakes Twitter users: real-time event detection by social sensors," in *Proceedings of the 19th international conference on World wide web*, 2010.

[5] I. Annamoradnejad and J. Habibi, "Top 50 trending topics (trends) of Twitter for 2018 (one hour interval)," *Mendeley Data, v1*, 2019.

[6] B. De Longueville, R. S. Smith, B. De Longueville, R. S. Smith, and G. Luraschi, "'OMG, from here, I can see the flames!': a use case of mining Location Based Social Networks to acquire spatiotemporal data on forest fires," *Lbsn*, 2009.

[7] K. Starbird, "(How) will the revolution be retweeted? Information diffusion and the 2011 Egyptian uprising," in *Proceedings of the ACM 2012 Conference on Computer Supported Cooperative Work*, 2012.

[8] G. Enli, "Twitter as arena for the authentic outsider: exploring the social media campaigns of Trump and Clinton in the 2016 US presidential election," *Eur. J. Commun.*, 2017.

[9] J. Nichols, J. Mahmud, and C. Drews, "Summarizing sporting events using twitter," in *Proceedings of the 2012 ACM international conference on Intelligent User Interfaces - IUI '12*, 2012.

[10] M. Nagarajan, H. Purohit, and A. Sheth, "A Qualitative Examination of Topical Tweet and Retweet Practices," in *International AAAI Conference on Weblogs and Social Media*, 2010.

[11] K. Lee, D. Palsetia, R. Narayanan, M. A. Patwary, A. Agrawal, and A. Choudhary, "Twitter Trending Topic Classification," pp. 251–258, 2011.

[12] D. Gao, W. Li, X. Cai, R. Zhang, and Y. Ouyang, "Sequential Summarization : A Full View of Twitter Trending Topics," vol. 22, no. 2, pp. 293–302, 2014.

[13] F. Liu, Y. Liu, and F. Weng, "Why is SXSW trending?: exploring multiple text sources for Twitter topic summarization," in *LSM '11 Proceedings of the Workshop on Languages in Social Media*, 2011.

[14] D. Chakrabarti and K. Punera, "Event Summarization using Tweets," *Proc. Fifth Int. AAAI Conf. Weblogs Soc. Media*, 2011.

[15] S. Lee and B. Kang, "Twitter Trending Topics Meaning Disambiguation," *... Acquis. Smart Syst. ...*, pp. 1–13, 2014.

[16] J. Hicap, "Guinness checking if #IVoteBTSBBMAs has set new record for most used hashtag in 24 hours," *Metro*, 2018.